\documentclass[prb,preprint]{revtex4-1}

\usepackage{amsmath}  
\usepackage{amsfonts} 
\usepackage{graphicx} 

\begin{document}

\title{Modeling Atomic Polarizability: From Charge Distortion to Non-Uniform Distributions}

\author{O. F. de Alcantara Bonfim}
\email{bonfim@up.edu} 
\affiliation{Department of Physics, University of Portland, Oregon, 97203}


\begin{abstract}
In the classical model of atomic polarizability, atomic charges are displaced by an applied electric field, assuming the electron cloud remains spherically symmetric but with its center shifted from the nucleus, thereby inducing an electric dipole. In this work, we propose that the applied electric field distorts the initially spherical electron cloud into a spheroidal shape, with the nucleus positioned at one of its focal points. We assume that the electron cloud's charge density remains uniform and is not altered by the applied field.
We derive a modified expression for the polarizability that accounts for the distortion factor, represented by the eccentricity of the spheroid, and analyze an additional model incorporating non-uniform charge distribution. Furthermore, we present an expression for polarizability that combines the effects of both distortion and non-uniform charge density. These models aim to refine the classical approximation and offer deeper insights into the mechanisms underlying atomic polarizability.
\end{abstract}

\maketitle

\section{Introduction}
Atomic polarizability is a fundamental concept in classical electrodynamics and quantum physics. It describes the tendency of an atom’s charge distribution to distort under an external electric field, inducing a dipole moment. This property is central to understanding dielectric behavior, van der Waals interactions, and the scattering and refraction of light. Atomic polarizability links microscopic atomic behavior to macroscopic material properties, making it indispensable in materials science, optics, and electronics \cite{Kwan}.

A neutral atom consists of a positively charged nucleus surrounded by a negatively charged electron cloud. In the absence of an external field, the charge distribution is spherically symmetric, with no net dipole moment. When an electric field is applied, the nucleus and electron cloud are displaced in opposite directions, inducing a dipole moment proportional to the applied field for weak fields. This relationship is expressed as:
\begin{equation}
\mathbf{p} = \alpha \mathbf{E},
\end{equation}
where $\mathbf{p}$ is the induced dipole moment, $\mathbf{E}$ is the external field, and $\alpha$ is the atomic polarizability. Factors such as atomic size, electron count, and effective nuclear charge influence $\alpha$. Larger atoms with loosely bound electrons are generally more polarizable than smaller ones with tightly bound electrons.

\section{A Simple Model}
A classical model \cite{Kwan, Griff} treats the electron cloud as a uniformly charged sphere and the nucleus as a point charge. When an external field displaces the nucleus and the electron cloud’s center by a distance $d$, the electron cloud remains spherically symmetric. The equilibrium configuration balances internal Coulomb forces and the external field.

Using Gauss’s law, the field at a distance $d$ from the center of a uniformly charged sphere is:
\begin{equation}
E = \frac{q d}{4\pi \varepsilon_0 a_0^3},
\end{equation}
where $q$ is the total charge, $a_0$ is the sphere’s radius, and $\varepsilon_0$ is the vacuum permittivity. Substituting $p = q d$ yields $\mathbf{p} = 4\pi \varepsilon_0 a_0^3 \mathbf{E}$, leading to:
\begin{equation}
\frac{\alpha}{4\pi \varepsilon_0} = a_0^3.
\end{equation}

For the hydrogen atom, semi-classical calculations \cite{Bow} give $\alpha / 4\pi \varepsilon_0 = \frac{21}{4}a_0^3$, while quantum mechanical calculations \cite{Zet} yield $\alpha / 4\pi \varepsilon_0 = \frac{9}{2}a_0^3$. Here $a_0$ is the Bohr radius. The simple model underestimates the quantum result by a factor of 4.5.

Attempts to improve the model include using the quantum mechanical charge density $\rho(r) = \frac{q}{\pi a_0^3}e^{-2r/a_0}$ and a rigid spherical cloud \cite{Griff}. However, calculations assuming $d \ll a_0$ yield $\alpha / 4\pi \varepsilon_0 = \frac{3}{4}a_0^3$, which is no improvement. Here, we propose a modified classical model that incorporates electron cloud distortion, as well as a model that assumes the electron cloud's charge distribution is non-uniform.

\section{A Model with Distortion}
In this model, the electron cloud retains a uniform charge density but distorts into a spheroid under the external field. The nucleus is at one focal point ($x = d$), and the external field is assumed to act in the positive $x$-direction. The boundary of the electron cloud is:
\begin{equation}
\frac{x^2}{a^2} + \frac{r^2}{b^2} = 1,
\end{equation}
where $r^2 = y^2 + z^2$, $b = a\sqrt{1 - \epsilon^2}$, and $\epsilon = d / a$ is the spheroid’s eccentricity.

To calculate the field at the nucleus, the electron cloud is conceptually divided into infinitesimal disks of radius $r$ and thickness $dx$, aligned perpendicular to the $x$-axis. The field from a disk at $x$ is:
\begin{equation}
dE_e = \frac{\rho}{2\varepsilon_0}\left[1 - \frac{|d - x|}{\sqrt{r^2 + (d - x)^2}}\right]dx,
\end{equation}
where $\rho = \frac{3q}{4\pi ab^2}$ and $r^2 = b^2\left(1 - \frac{x^2}{a^2}\right)$. Simplifying and integrating over $(-a, a)$, the net electric field at $x = d$ is:
\begin{equation}
\label{dfield}
E_e = \frac{\rho d}{2\varepsilon_0}g(\epsilon),
\end{equation}
where $g(\epsilon)$ is:
\begin{equation}
g(\epsilon) = 2 - \frac{2}{\epsilon^2} + \frac{1}{\epsilon}\left(1 - \frac{1}{\epsilon^2}\right)\ln\left(\frac{1 - \epsilon}{1 + \epsilon}\right).
\end{equation}
For $\epsilon \ll 1$, $g(\epsilon) \approx \frac{2}{3}\left(1 - \frac{2}{5}\epsilon^2\right)$.
We set $ab^2 = a_0^3$ where $a_0$ is the atomic radius, as the electron cloud density is assumed to remain the same after the atom polarizes.
Equating the applied and electron cloud fields gives:
\begin{equation}
\frac{\alpha}{4\pi \varepsilon_0} = \frac{2}{3g(\epsilon)}a_0^3.
\end{equation}
As expected, for $\epsilon \to 0$, $g(0) = 2/3$, and the classical result for the polarizability with no distortion is recovered. The modified expression for atomic polarizability with distortion offers an improvement over the simple model. However, significant distortion ($\epsilon \approx 0.97$) is required to match quantum mechanical results, which may not represent a physically realistic distortion. The spheroid model is a generalization of the simple model, taking into account a more realistic response of the electron cloud to the external field.

\section{A Model with Non-Uniform Charge Distribution}

Another model for polarizability can be realized by assuming that the electron charge distribution consists of two parts: the first is a uniform distribution over the entire volume of the atom $(\rho_V)$, and the second $(\rho_N)$ is also uniform but superposed on the first, with a higher charge density enveloping the nucleus. These charge distributions are assumed to remain unchanged after the application of an external electric field, as in the simple model. From charge conservation, the total charge of the electron cloud is:
\begin{equation}
\label{charge}
q = \rho_V \frac{4}{3}\pi a_0^3 + \rho_N \frac{4}{3}\pi R_N^3,
\end{equation}
where $\rho_N = \beta\rho_V$ and $R_N = \gamma a_0$, with $\beta > 1$ and $\gamma < 1$. $R_N$ represents the radius of the higher charge density ($\rho_N$) region of the electron cloud around the nucleus. The electric field at the nucleus, when it moves to $x = d$ under the influence of the external electric field, is given by:
\begin{equation}
E_e = \frac{\rho_V d}{3\varepsilon_0}.
\end{equation}

Imposing $E = E_e$, and using Eq.~\eqref{charge} along with the definition of polarizability, we obtain:
\begin{equation}
\frac{\alpha}{4\pi\varepsilon_0} = (1 + \beta\gamma^3)a_0^3.
\end{equation}

This expression improves upon the simple model. By choosing $\beta = 8$ and $\gamma = 0.76$, we find $\alpha / 4\pi\varepsilon_0 = 4.5a_0^3$, which reproduces the quantum mechanical result for the hydrogen atom, though the parameters used may not be entirely realistic. Comparing the two models introduced here, we find that the model with a non-uniform charge distribution provides a larger contribution to polarizability than the model based on charge distortion.
Each model, however, independently improves upon the simple model presented in the literature \cite{Kwan, Griff}.
While atomic polarization is fundamentally a quantum mechanical phenomenon, classical models provide valuable insights despite their inability to fully replicate quantum mechanical calculations of polarizability. Nevertheless, these models can still serve as useful tools in elucidating the most significant mechanisms driving the polarization process.

\section{Combined Model}

When both mechanisms are present, we can combine Eq. (\ref{dfield}) (with $\rho$ replaced by  $\rho_V$) and Eq. (\ref{charge}). This results in the expression for polarizability when both distortion and non-uniform charge effects are considered:
\begin{equation}
\frac{\alpha}{4\pi\varepsilon_0} = (1 + \beta\gamma^3)\frac{2}{3g(\epsilon)}a_0^3.
\end{equation}
This combined model captures the fundamental mechanisms behind the induction of an electric dipole in an atom. Any further refinements would be quantitative in nature, involving adjustments to the shape and charge distribution of the electron cloud in response to the external electric field.

\section{Conclusion}

We proposed a classical model for atomic polarizability in which the electron cloud distorts into a spheroid under an external field while maintaining a uniform charge density. Additionally, we derived an alternative model incorporating non-uniform charge distribution within the electron cloud.  While the distortion model captures the effects of geometry, the non-uniform charge model achieves closer agreement with quantum mechanical results. Combining these effects offers a more comprehensive classical description of atomic polarizability.
These models provide accessible approaches for teaching polarizability in introductory electrodynamics courses.

\begin{acknowledgments}
The author thanks David Griffiths for reading the manuscript and providing valuable suggestions. 
\end{acknowledgments}



\end{document}